# SPECI, a simulation tool exploring cloud-scale data centres

Ilango Sriram[1]

[1] University of Bristol, Department of Computer Science,
Merchant Venturers Building, Woodland Road, Bristol BS81UB, United Kingdom
ilango@cs.bris.ac.uk

**Abstract.** There is a rapid increase in the size of data centres (DCs) used to provide cloud computing services. It is commonly agreed that not all properties in the middleware that manages DCs will scale linearly with the number of components. Further, "normal failure" complicates the assessment of the performance of a DC. However, unlike in other engineering domains, there are no well established tools that allow the prediction of the performance and behaviour of future generations of DCs. SPECI, Simulation Program for Elastic Cloud Infrastructures, is a simulation tool which allows exploration of aspects of scaling as well as performance properties of future DCs.

**Keywords:** Cloud computing, data centre, middleware, scaling of performance, simulation tools.

## 1 Introduction

The current trend towards cloud computing drives the demand for IT resources provisioned from data centres (DCs). Economies of scale accelerate the growth in the size of DCs, and over the coming decades are very likely to lead to DCs several orders of magnitude larger than the biggest ones today. IBM's project Kittyhawk already has the vision of one day building a DC for high performance computing that is big enough to host a single application as big and powerful as today's entire internet [1].

However, not all properties are expected to scale linearly when adding more components and increasing the scale of DCs. Further, the number of components will be so large as to impose "normal failure". For example with 500,000 servers in the DC, if the average life expectancy of a server is three years and the time to a temporary failure a few months [2], then on average 650 servers will need to be replaced every day with multiple temporary failures every minute. The state of near permanent hardware failures has to be taken into account by the DC's resilience mechanisms. Furthermore, load-balancing of virtual machines will cause continuous dynamics in the system. These constant changes have to be taken into consideration when assessing the overall performance capabilities of a data centre.

In most engineering fields there are predictive tools that allow simulation of engineered products that are not yet built. While in other domains these can model the following generations of technology with high precision given the computing resources, such as SPICE for circuits on microchips [3], there is only limited understanding, and there are no well-established predictive tools for data centres. We propose SPECI, Simulation Program for Elastic Cloud Infrastructures, a simulation tool that enables exploration of scaling properties of large data centres. The aim of this project is to simulate the performance and behaviour of data centres, given the size and middleware design policy as input.

The rest of this paper is organised as follows. Section 2 discusses cloud computing and its current state which forms the motivation for this work and introduces middleware of data centres. Section 3 introduces the related work in the field of cloud computing, and introduces Simkit, the simulation framework that was used. Section 4 explains how scalable middleware will be executed at the node level, and that the method used to communicate with the nodes will partially determine the performance overhead of the data centre: in this section a simplified problem of communicating aliveness of components for the resilience component is introduced; this builds the basis for SPECI. Section 5 explains the current architecture of SPECI. A case study with SPECI was conducted and is presented in Section 6, before Section 7 closes with discussion of planned future work and conclusion.

## 2. Background

### 2.1 Cloud Computing

Cloud computing is a growing form of IT provision, where resources are delivered as a service from data centres. The advent of large-scale, commodity-computer data centres built at low-cost locations is a driver of this technology.

There has been a long debate about what cloud computing stands for. This is coming to an end, and analysts' position papers have converging definitions, such as Gartner [4], Forrester Research [5], and The 451Group [6]. Recently, further, a widely accepted working definition of Cloud Computing has been published by the American NIST institute [7]. In general an IT provision is cloud computing, if it is (1) delivered as a service (2) using Internet technologies from (3) massively scalable (elastic) resources which are generally (4) shared with other customers (multi-tenancy), where (5) resources are distributed dynamically and redistributed on demand. In addition to this, cloud computing is an economic model of billing resources (6) by consumption in metered quality and quantity (pay-as-you-go).

There is a differentiation between public clouds and private clouds. While public clouds are run by utility providers, private clouds are owned, operated, and under control of the company that uses them. Private clouds nevertheless use the same service technology and offer interoperability. This allows elasticity into the public cloud which is called cloudbursting, so that consumers can use public clouds if they need additional resources.

In practice, cloud computing is delivered in three forms: Infrastructure as a Service (IaaS), Platform as a Service (PaaS) and Software as a Service (SaaS). IaaS provides bare resources or Virtual Machines (VMs). PaaS goes beyond the bare provision of resources, and gives consumers a preconfigured platform for their work. With PaaS, developers can develop the software they are working on, but do not have to care about requirements such as an SDK or a database, or a framework for scaling the application to use further resources. These requirements are provided by the platform. Finally with SaaS, ready to use hosted software solutions are provided to consumers.

Industry analysts are in consent that the development of mature cloud computing is well underway. Although cloud computing is not likely to completely replace traditional IT in the near future, for some application areas it is set to replace current IT delivery from in-house DCs [9]. However precise the figures are, the turn towards cloud computing is obvious, and it will accelerate the demand for DCs of increasing scale. The more cloud computing gets adopted and standardised, the less qualitative differences between offers will exist, and the more provisions will be differentiated purely on cost. To further cut costs, economies of scale will bring in a demand for even larger DCs. As mentioned above, however, not all properties in DCs and its middleware will scale linearly and it is not known how the behaviours of following generations of DCs are going to be. There is lack of predictive simulation tools; this is something SPECI is addressing.

## 2.2 Normal Failure and Middleware Scalability

Cloud data centres are built using commodity hardware, and can be divided into racks, blades, and VMs. Unlike in high performance computing (HPC) where DCs are often custom-built for the purpose of the computations and a significant part of computing power is gained from graphical processing units (GPUs), cloud DCs rely on inexpensive traditional architecture with the key components being CPUs, memory, discs, and network. As economies of scale are driving the growth of these DCs, the sheer number of off-the-shelf components used in coming decades in combination with each component's average life cycle will imply that failure will occur continually and not just in exceptional or unusual cases. This expected near-permanent failing of components is called "normal failure". For cost reasons, the DC operator will leave the failed components in place and from time to time replace the blades on which failure occurred or even entire racks on which several blades have failed. The impact of failure and resilience or recovery needs to be taken into account in the overall performance assessment of the system.

The components of the DCs are tethered by a software layer called middleware, which takes care of job scheduling, load-balancing, security, virtual networks, and resilience. It combines the parts of the DC together and is the management layer of the DC. When the numbers of components in the DC increases, the middleware has to handle it. It is unlikely that all properties in middleware will scale linearly when scaling up the size of DCs. Currently, there is a lack of predictive tools or methods to estimate the performance of the middleware and thus of the DC before building it.

Therefore, there is a need for simulation tools that can help us evaluate the behaviour and performance with reproducible results, prior to designing and building such DCs. In the absence of such simulation tools one has to rely on theoretical simplified models or build the system and simply hope it performs well. The latter is undesirable and imposes financial drawbacks, if systems can't be tuned before developing and deploying.

## 3. Related Work

### 3.1 Cloud Computing

So far there are only a few scientific publications on technologies that are enabled by cloud computing, such as CloneCloud [10] which enables resource intensive computations like face recognition on smartphones by automatically and transparently cloning the context of the computations into the cloud. We expect to see an increase in academic publications describing new technologies, but at the moment the majority of publications in cloud computing are either management articles or come from practitioners of utility computing and grid computing.

In the area of performance assessment of cloud DCs, there is some preliminary work going on within the Open Cirrus project [11]. They have built a cloud research test bed running Eucalyptus [12], which is an open source implementation of the EC2 interface. However, so far they are only looking at the performance of individual virtual machines in cloud environments at Amazon EC2 in comparison to execution in local DCs, and not at the performance capabilities of the entire DCs.

Vishwanath et al. [18] have looked into performance and cost for datacenters that consist of modularized shipping containers which are not serviced for hardware faults until the entire container gets replaced. Further, the HP Cells as a Service project [13] is developing a prototype middleware and management system for cloud infrastructures that is scalable, reliable and secure. While it achieves security using virtual machines, virtual storage volumes and virtual networks, details of how they solve reliability and tolerate the continuous failures that occur in large-scale DCs, and how they solve scalability performance issues, are not yet public.

There is no known work so far on predicting scaling issues for future generations of commodity-computer DCs. However, there is CloudSim [14], a simulation framework to evaluate the performance of cloud computing infrastructure. The simulator is built on top of a grid computing simulator (GridSim) and looks at the scheduling of the execution application, and the impact of virtualisation on the application's performance. However, in this project our interest is more on the DC provider side. We assume that the cloud elasticity is big enough to not reach bottlenecks in the execution of applications, but we do want to know how the overall DC and in particular the middleware that tethers the network of virtual services can perform with increasing numbers of components. Further, we believe that grid architecture and virtualisation technique used for cloud computing are two competing technologies that will not be used in combination. Running a grid under the virtualisation layer adds significant complexity without offering any obvious advantage.

### 3.2 Simulation method: Simkit

In the absence of real test beds, of alternative physical representations, or of precise formal models, simulation helps in exploring assumptions about models before building the systems. Discrete event simulations (DES) [15] are a type of simulation where events are ordered in time, maintained in a queue of events by the simulator, and each processed at given simulation time. This means the model is time based, and takes into account resources, constraints and interactions between the events as time passes. Central to DES are a clock and an event list that tells what steps have to be executed. In order not to re-implement common features of DES, SPECI uses an existing package for DES in Java. There exist several such packages and toolkits, and we chose SimKit [16] which was one of few Java packages that were updated recently. It implements the clock using a queue of events, each of which is associated with a start time. The computation of the event then takes place with duration of zero time interval. When the computation of the event has finished, the clock advances to the time of the next event in the schedule. Simkit also offers many distributions for random-number generation.

Simulation tools are common in other domains: For example in the microelectronics industry there is the circuit simulator SPICE [3], that allows one to simulate the behaviour of future designs of chips with high precision before actually building them, given the computing resources. With the help of this simulation tool better chip designs can be found, and verified quicker and at lower cost. Similarly, SPECI is intended to give us insights into the expected performance of DCs when they are designed, and well before they are built.

## 4. SPECI Example: Scalable Middleware

DCs are managed by middleware which provides functionality such as job scheduling, load-balancing, security, virtual networks, and resilience. Because many of these settings change very frequently, it needs to continuously communicate new policies to the nodes. Scalable middleware can either manage its constituent nodes using central control nodes, which is a poorly scaling hierarchical design, or it can manage the DC using policies, which are broken into components that can be distributed using peer-to-peer (P2P) communication channels and executed locally at each node. This better scalable solution can cause a problem of timeliness of how quickly updated policies will be available at every node, and of consistency whether the same policies are available and in place everywhere. A certain overhead load for the management will be generated in either case, which will determine the performance loss when scaling the DC by adding more components. [17]

As a first step, we have built a simulator to observe the behaviour of part of the middleware that recognises failed components across the network of systems. This failure communication mechanism can be seen as a simplified substitution for the policy distribution problem.

We were interested in the behaviour of a system with a large number of components, where each component can be working correctly or exhibiting a temporary or permanent failure. Failures occur frequently in large DCs given the number of components and the expected lifetime of each of them. Any one component cooperates with some of the other components, is thus interested in the aliveness of these and performs queries to find this out. As the number of components increases, the number of states that have to be communicated over the network increases. We need to know what happens with our system in terms of how well in time can the states be communicated and at the cost of what load. This setup is of interest to any computing facility with such a large number of components where some will be near permanently failing or other changes need to be communicated frequently. To find out how various protocols may scale, and how quickly or whether at all a consistent view of the state of cooperating nodes can be achieved under certain conditions, a set of simulation experiments was set up, as described in the following paragraphs.

There is a number (n) of nodes or services connected through a network. Each of these nodes can be functioning (alive) or not (dead). To discover the aliveness of other nodes, each node provides an arbitrary state to which other nodes can listen. When the state can be retrieved the node is alive, otherwise it is dead. The retrieval of aliveness of other components is called "heartbeat". Every node is interested in the aliveness of some of the other nodes, the amount of "some" being configurable. Each node maintains a subscription list of nodes in whose aliveness it is interested. We are interested in how the implementation of the heartbeat retrieval affects the system under given configurations, when the total number of nodes n increases.

Several architectures of heartbeat retrieval could be possible. First, there could be central nodes that collect the aliveness of all other nodes and then inform any node interested in any particular state. Second, there could be a hierarchical design where depending on the number of hierarchy levels certain nodes would gather the information of some other nodes, and make them available to their members and to the node next higher in the hierarchy. Third, there could be a simple P2P mechanism where any node simply contacts the node of interest directly. Then, there could be a smarter P2P protocol where a contacted node would automatically reply with all aliveness information it has available of other relevant nodes.

The investigation reported here was set up to observe the behaviour of the overall system under these protocols and various change rates when the number of nodes involved scales up. The simulations address a number of questions. The first question of interest is, what the overall network load is for each of the above protocols under given settings and size, and how much data has to be sent over the network in a given time period. Second, there is significant interest in how the "time-for-consistency" curve of the system looks like. This means, after simultaneous failure or recovery of a number of nodes, after how many time-steps changes are propagated through the entire system, and if there are continuous failures appearing, how many nodes have a consistent view of the system over time? It is of further interest to see how many time-steps and how much load it takes until new or recovered nodes have a consistent view of the system, and how many time-steps it takes to recover after failure of a large number n of nodes, or for recovery of the entire network. There is also interest in the trade-off between timeliness and load for each of the protocols in the sense of how much extra load will be required to retrieve a better or more consistent view. In other words, for how much load can one get what degree of timeliness?

## 5. Simulator Architecture

The implementation of SPECI is split in two packages, one represents the data centre layout and topology, and the other one contains the components for experiment execution and measuring.

The experiment part of the simulator builds upon SimKit, which offers event scheduling as well as random distribution drawing. SimKit has preconfigured pseudo random classes for common distributions, which return the same value for repeated executions. This makes it possible to execute repeated runs with modified settings, and in each run to receive the same random draws, and thus the same scheduling sequence and scheduling times for events.

The simulation entry class is a wrapper class that contains the configurations of all runs. It triggers the Simkit engine to start the simulations and handles the statistical analysis of the output once the runs have terminated. The Simkit engine always starts simulations by calling the method doRun() of all existing objects in the project where implemented. These are used to start the experiments, and need to trigger accordingly configured parameter change listeners or place new events on the scheduling engine. In this simulator, there is only one doRun() method in the singleton handler. This method creates the DC setup from the data centre layout package with the specifications provided. It then adds three types of events to the event scheduler: probing events, update events, and failure events. The first probing event is generated at 2.0 seconds simulation time to allow instantiation before measuring potential inconsistencies in the system. When this event is triggered, all subscriptions are tested for inconsistencies against the real state and the total number passed on to a class for collecting tally statistics of the simulation model. Before the probing event terminates, it reschedules itself for the next execution 1.0 seconds later than the current simulation time. Thus, every second a monitoring probe is passed on for an evaluation after termination of the simulation. Further, the handler generates one update event for every node in the data centre. This event triggers the node to update the list of its subscriptions. These heartbeat retrieval events are drawn from a uniform distribution with a delay between 0.8 and 1.2 seconds and reschedule themselves with a delay from the same distribution. Similarly, the handler schedules the occurrence of the first failure. The time to the next failure is variable in our experiments and has to be specified in form of a parameterised random function. When the failure event is triggered, it picks a node at random which it will set to have failed. If the failure function picks a node that is already failed, it will act as repair event and bring the component back alive. Alternatively, failed components are not repaired, and kept until the entire shipping container is replaced, as proposed in Vishwanath's [18] DC model.

The data centre layout package contains classes for each type of component in the data centre, such as nodes and network links. These components mimic the operations of interest in the observed data centre, such as the transfer of network packets, maintaining subscriptions to other nodes, and keeping subscriptions up to date using the policy chosen for the experiment. The components have monitoring points that can be activated as required by the experiment. As simplification the network topology assumes a one hop switch, as this work is not interested in routing and the load on parts of the network, but rather on individual network links associated to a node and the entire network. The data centre package further contains a component that maintains a global view of the data centre to deal with the connection and referral logic, which is only used when the topology chosen is a centralised heartbeat retrieval or policy distribution, such as the central or hierarchical one. In the central case this is a list of providers, which pass on information to all other nodes. In the hierarchical case, the global view knows of the hierarchy levels, and which node ought to request information from which other node, as described in Section 4. If the setup configuration uses the simple P2P or transitive P2P communication channel, then the communication logic is dealt by the nodes, as in this case only a local view of the system is required. Depending on the used policy, some, none or all nodes can act as providers and pass on information they have about other nodes. In reality this passing on can cause delayed information, as the information stored and passed on is not necessarily real time. In this simulator there is a configurable threshold of say one second, which is the maximum permitted age information can have to still be passed on. If the information is older, the providing node will not pass on this data, but instead retrieve newer data by the respective mechanism. If nodes are provider nodes, they have the option to only accept a maximum number of requests per time interval.

In the initialisation phase at runtime, the simulator creates an object for each node and network link in the data centre, subscribes all nodes to some other ones with a distribution as specified in the configuration, and loads the communication policy for the setup. The rest of the runtime is entirely driven by the event queue. The model terminates when the specified simulation time has expired. The simulator will then calculate statistics collected by tally statistics classes. Further more detailed monitoring data is written to files. Therefore, while each object is retrieving the heartbeat of its subscriptions, the load generated is monitored, and aggregated access counts per component over a configurable duration stored to a file. Similarly, when a failure occurs, the time and the number of nodes which have become inconsistent with the actual state of the landscape gets saved to another file. After the simulations are executed these data files can be visualised independent of the simulator.

## 6. Case Study

In this section, we present a case study made using SPECI in which we observe the number of nodes that have an inconsistent view of the system. This is the case if any of the subscriptions a node has contains incorrect aliveness information. We measure the number of inconsistencies by probing the count every second. After an individual failure occurs, there are as many inconsistencies as there are nodes subscribed to the failed node. Some of these will regain a consistent view before the following observation, and the remaining ones will be counted as inconsistent at this observation point. If the recovery is quicker than the time to the next failure at the following observations less nodes will be inconsistent until the curve drops to zero, and the inconsistency curve could look like Figure 1. This probing was carried out while running SPECI with increasing failure rates and scale. Runs were carried out for DC sizes of $10^2$, $10^3$, $10^4$, and $10^5$ nodes. Assuming the number of subscriptions grow slower than the number of nodes in a DC, we set the number of subscriptions fixed to the square root of the number of nodes.

For each of these sizes a failure distribution was chosen such that on average in every minute 0.01%, 0.1%, 1%, and 10% of the nodes would fail. Because this work is essentially exploratory, a gamma distribution and a pair of coefficients that would result in the desired number of failures were picked. For each pair of configurations 10 runs, each lasting 3600 simulation time seconds, were carried out and the average number of inconsistencies along with its standard deviation, maximum, and minimum number were observed. The half width of the 95% confidence intervals (95% CI) was then calculated using the Student's t-distribution for small or incomplete data sets.

Figure 2 shows that the average number of inconsistencies increases when the failure rate increases, and also when the number of nodes increases. Figure 3 shows the same data as Figure 2, but the first few data points are plotted on a linear scale. This makes the confidence intervals visible, and one can see for small DCs or small failure rates the two protocols differ insignificantly. But as these numbers increase, the mean of one protocol moves out of the other protocol's confidence interval, and when the sizes get bigger the confidence intervals get distinct as can be seen for 1000 nodes and 1% failure rate. This shows that with growing size and failure rates, the choice of the protocol becomes more significant, and also that the P2P protocol scales better for the objective of low inconsistencies under the given simplifications. The surprising result here is, given there were identical polling intervals, that the transitive P2P was expected to be the protocol with the biggest delay due to the fact that delays would accumulate with forwarding. However, for such an accumulating of age for aliveness data to be observable it is necessary to have larger numbers of subscriptions or to generate the subscriptions with a structure so that the chance of subscribing to a node is higher if the node has similar subscriptions, because only then enough transitive subscription sharing is available. Figure 4 shows inconsistencies grouped by failure rates. To compare the values of different DC sizes, the number of inconsistencies is normalised by the number of nodes. At the same time, the number of inconsistencies still grows with the size of the DC. This suggests that none of the protocols scale linearly. On the other hand, when grouping by DC sizes, these normalised values increase by one order of magnitude when the failure rate increases by such. This suggests that the failure tolerance of both protocols appears robust.

The careful reader might have noticed that we were interested in performance drawbacks when the size of the data centre increases, but in this paper we focussed on the inconsistencies under each of the protocols. The number of inconsistencies can be reduced by reducing the polling interval of each of the nodes at the cost of additional load. Analysing the performance of the protocols shall be left for future work.

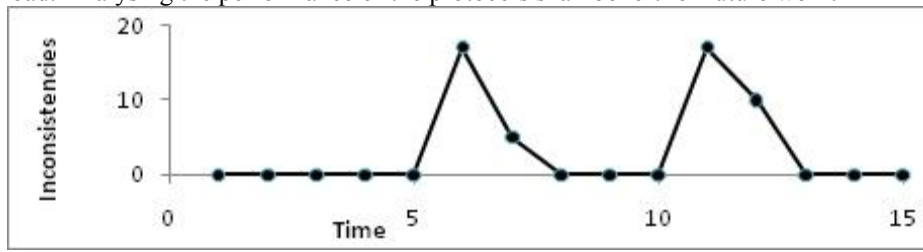

**Figure 1: Inconsistency probes during recovery from failures**

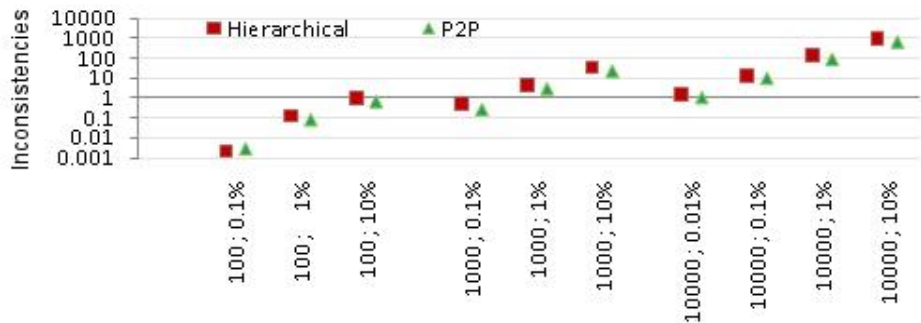

**Figure 2: The mean of Inconsistencies increases with the failure rate**

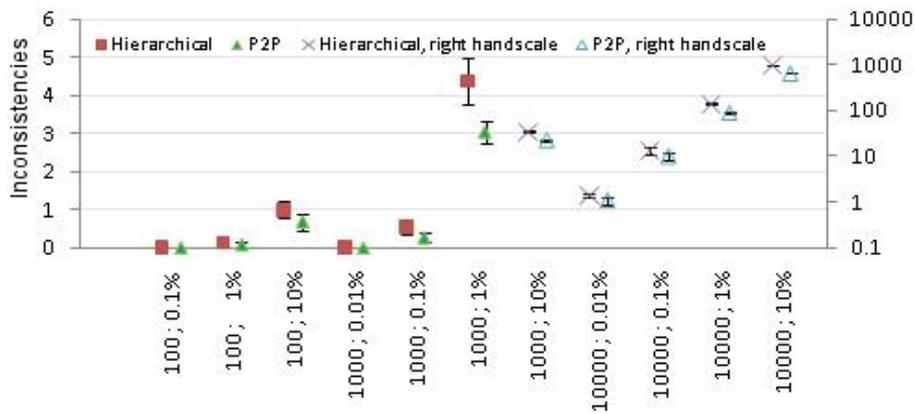

**Figure 3: Mean of Inconsistencies and their confidence intervals, linear vertical scale for smaller and logarithmic scale for larger values. With increasing size differences become significant.**

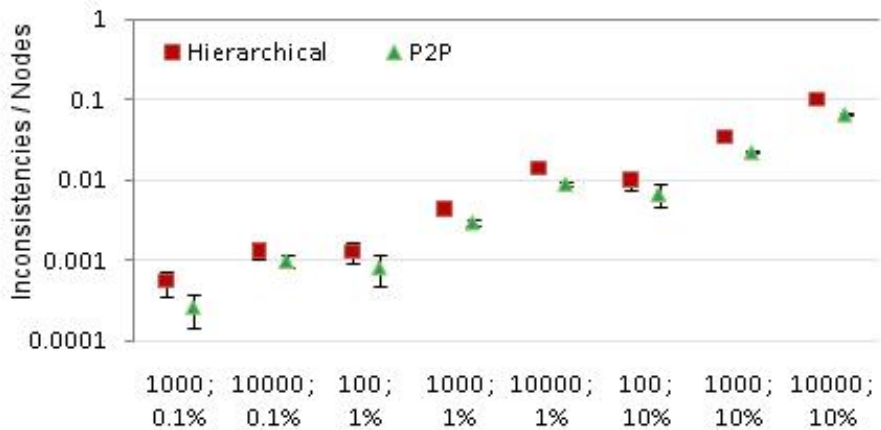

**Figure 4: Inconsistencies normalized by number of nodes. Both protocols scale linearly with the failure rates, but not with the number of nodes.**

## 7. Future work

Imminently, further case studies with SPECI are planned. First, we are interested in using other failure and recovery mechanisms. This includes matching failure rates to those in literature and using recovery mechanisms where failed nodes are not replaced until the entire building unit gets replaced. Second, runs for larger DC sizes are to be carried out. In addition it should be observed how the system behaves when using different numbers of subscriptions. In combination with the transitive P2P this could show whether any benefits or differences from node cliques can be found. Further, correlated failure and conditions where a huge amount of nodes or the entire DC fail at the same time need to be simulated. Then, it is necessary to combine the simulation's load measurements with measurements of inconsistencies during varying failure rates. This can show what the failure and load thresholds which prevent the system from ever reaching a consistent status are, and what settings make it impossible for the system to recover. These simulations can also be used to suggest a load and consistency trade-off for middleware mechanisms.

The following step is to look for alternative models to verify the findings. These could be mathematical or formal models, and for smaller DCs comparison with values that can be measured. For medium term future work, it is necessary to expand SPECI. At the moment it looks at one dimensional state communication problems. Real middleware has to distribute several policies over the network. It needs to account for VMs and load-balancing, security, and job scheduling. SPECI must become capable of modelling such multidimensional problems that middleware is facing in order to access the scaling properties of future cloud-scale DCs.

## 8. Conclusion

When designing scalable middleware, centralised orchestration will not be feasible; instead it will be necessary to have the system orchestrate itself with just the given local view and without knowledge of the entire system. Even then, it is expected that DCs do not scale linearly when they get larger and contain more components. Practitioners need to know about the scaling properties before building these DCs. In this paper we have presented SPECI, a simulation tool which allows exploration of aspects of scaling as well as performance properties of future DCs. SPECI was then used to look at inconsistencies that arise after failures occur, and it could be shown at the example of the communication of failures, that when the size and failure rate of the DC increases, a distributed DC management becomes favourable.

## A. Acknowledgements

This project is part of a PhD funded by Hewlett-Packard Labs' Automated Infrastructure Lab. We thank Hewlett-Packard for the interest in this topic and support.

# References

1. Appavoo, J., Volkmar, U. and Waterland, A.: Project Kittyhawk: building a global-scale computer: Blue Gene/P as a generic computing platform. SIGOPS Oper. Syst. Rev. 42(1), 77-84, (2008)
2. Failure Rates in Google Data Centers. http://www.datacenterknowledge.com/archives/2008/05/30/failure-rates-in-google-data-centers/
3. Nagel, L.W.: SPICE2: A Computer Program to Simulate Semiconductor Circuits. Technical Report No. ERL-M520, University of California, Berkeley, (1975)
4. Cloud, SaaS, Hosting, and Other Off-Premises Computing Models. Gartner Research. (2008)
5. Gillett, F.E.: The New Tech Ecosystems Of Cloud, Cloud Services, And Cloud Computing. Forrester Research. (2008)
6. The 451 Group: Partly Cloudy: Blue Sky Thinking about Cloud Computing. (2008)
7. Mell, P., Grance, T.: Draft NIST Working Definition of Cloud Computing. National Institute of Standards and Technology, Information Technology Laboratory (2009)
8. Clearwater, S.H. and B.A. Huberman, Swing Options: a Mechanism for Pricing IT Peak Demand. In: International Conference on Computing in Economics. (2005)
9. Gillett, F.E.: There are Three IT Architectures, Not One. Forrester Research. (2007)
10. Chun, B., and Maniatis, P.: Augmented Smart Phone Applications Through Cloud Clone Execution. In: HotOS XII (2009)
11. Baun, C. et al.: Elastic Cloud Computing Infrastructures in the Open Cirrus Testbed Implemented via Eucalyptus. In: ISGC 2009. Forthcoming 2009 LNCS, Springer, Heidelberg http://bit.ly/2Ck3tv
12. Nurmi, D., Wolski, R., Grzegorczyk, C., Obertelli, G., Soman, S., Youseff, L., and Zagorodnov, D.: The eucalyptus open-source cloud-computing system, In: CCGrid, pp.124-131, 9th IEEE/ACM International Symposium on Cluster Computing and the Grid, (2009)
13. HP Labs: Cells as a Service. http://www.hpl.hp.com/open_innovation/cloud_collaboration/projects.html
14. Buyya, R, Ranjan, R., and Calheiros, R. N.: Modeling and Simulation of Scalable Cloud Computing Environments and the CloudSim Toolkit: Challenges and Opportunities. In: Proceedings of the 7th High Performance Computing and Simulation (HPCS 2009) Conference, Leibzig, Germany, (2009)
15. Ferscha, A.: Parallel and Distributed Simulation of Discrete Event Systems. In: Parallel and Distributed Computing Handbook, ed. A. Y. Zomaya, pp. 1003-1041. McGraw-Hill, New York (1996)
16. Buss, A.: Simkit: Component based simulation modeling with Simkit. In: 34th Conference on Winter Simulation, pp. 243-249. (2002)
17. Isard, M.: Autopilot: automatic data center management. *SIGOPS Oper. Syst. Rev.* 41, 2 (2007)
18. Vishwanath, K. V., Greenberg, A., and Reed, D. A: Modular data centers: how to design them?. In: Proceedings of the 1st ACM Workshop on LSAP (2009).